\documentclass[sigconf]{acmart}
\usepackage{multirow}
\usepackage{amsmath}
\AtBeginDocument{%
  }

\setcopyright{acmlicensed}
\copyrightyear{2025}
\acmYear{2025}
\acmDOI{XXXXXXX.XXXXXXX}
\acmConference[AHs '25]{International Conference on Augmented Humans 2025}{March 16--20}{Ab Dhabi, UAE}
\acmISBN{978-1-4503-XXXX-X/2018/06}

\begin{document}

\title{Conditions for Inter-brain Synchronization in Remote Communication: Investigating the Role of Transmission Delay}

\author{Sinyu Lai}
\affiliation{%
  \institution{The University of Tokyo}
  \city{Tokyo}
  \country{Japan}
}
\email{sinyu-lai@g.ecc.u-tokyo.ac.jp}

\author{Wanhui Li}
\affiliation{%
  \institution{The University of Tokyo}
  \city{Tokyo}
  \country{Japan}
}
\email{li-wanhui409@g.ecc.u-tokyo.ac.jp}

\author{Kaoru Amano}
\affiliation{%
  \institution{The University of Tokyo}
  \city{Tokyo}
  \country{Japan}
}
\email{kaoru_amano@ipc.i.u-tokyo.ac.jp}

\author{Jun Rekimoto}
\affiliation{%
  \institution{The University of Tokyo}
  \institution{Sony CSL - Kyoto}
  \city{Tokyo, Kyoto}
  \country{Japan}
}
\email{rekimoto@acm.org}


\begin{abstract}
Inter-brain synchronization (IBS), the alignment of neural activities between individuals, is a fundamental mechanism underlying effective social interactions and communication. Prior research has demonstrated that IBS can occur during collaborative tasks and is deeply connected to communication effectiveness. Building on these findings, recent investigations reveal that IBS happens during remote interactions, implying that brain activities between individuals can synchronize despite latency and physical separation. However, the conditions under which this synchronization occurs or is disrupted in remote settings, especially the effect of latency, are not fully understood. This study investigates how varying transmission latency affects IBS, in order to identify thresholds where synchronization is disrupted. Using electroencephalography measurements quantified through Phase Locking Value—a metric that captures synchronization between brainwave phases—we first confirm synchronization under face-to-face conditions and then observe changes in IBS across remote communication scenarios. Our findings reveal that IBS can occur during remote collaboration, but is critically dependent on transmission delays, with delays exceeding 450 ms significantly disrupting synchronization. These findings suggest that IBS may serve as a key indicator of communication quality in remote interactions, offering insights for improving remote communication systems and collaboration.
\end{abstract}



\begin{teaserfigure}
  \includegraphics[width=\textwidth]{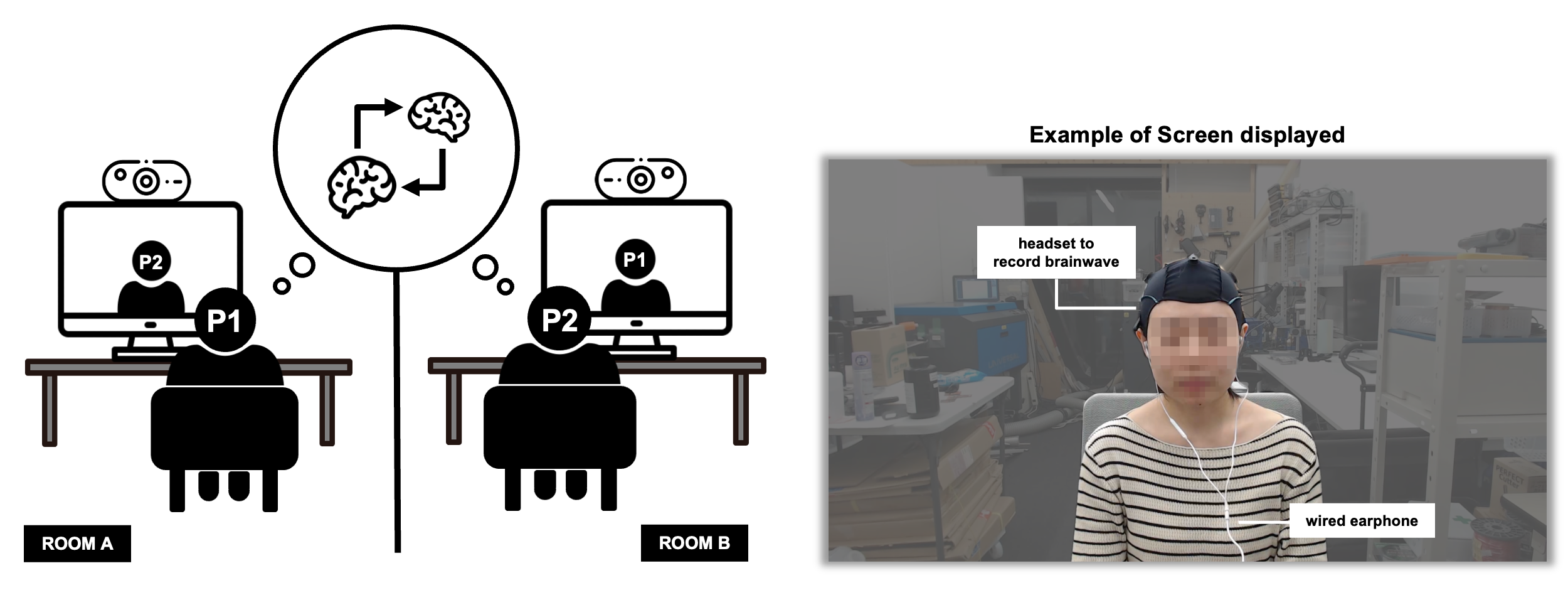}
  \caption{Brainwave activity is measured during remote communication with varying transmission delays to observe differences in inter-brain synchronization. Participants wear headsets to measure brainwaves and use wired earphones during the communication.}
  \label{fig:teaser}
\end{teaserfigure}


\maketitle

\section{INTRODUCTION}
Remote communication has become increasingly prevalent, driven by advancements in technology and the growing need for flexibility in work and social interactions \cite{raghuram2019virtual}. Despite its widespread adoption, significant differences in communication quality exist between face-to-face and remote interactions. Face-to-face communication is particularly effective in fostering cooperation, building group identity, and coordinating complex tasks—benefits that can be more challenging to achieve in remote settings \cite{6287790}. One key reason is that face-to-face interactions allow for immediate feedback and the seamless exchange of non-verbal cues, both of which contribute to a richer quality of interaction \cite{argyle2013bodily} \cite{pentland2012new}. In contrast, remote communication often leads to ``nonverbal overload,'' where participants must consciously interpret facial expressions, gestures, and eye contact, processes that would normally be automatic in face-to-face settings \cite{Bailenson2021Nonverbal}. These distinctions highlight the importance of understanding and addressing the unique challenges of remote communication to optimize collaboration and minimize potential drawbacks.

Over the past few years, numerous studies have utilized hyperscanning, a technique that measures the brain activity of two or more individuals at the same time, to explore the neural mechanisms underlying social interaction \cite{app10196669}. The development of hyperscanning technology has enabled researchers to observe the simultaneous brain activity of multiple individuals, leading to the discovery of Inter-Brain Synchronization (IBS) \cite{10.1371/journal.pone.0012166}. IBS refers to the alignment of brainwave activities (electrical signals generated by neurons) between individuals during communication and is commonly calculated by analyzing the coherence or phase-locking value (PLV) between the neural oscillations of individuals \cite{KELSEN20221249}. This phenomenon has been extensively documented in face-to-face interactions, where synchronized brain activity has been linked to successful communication outcomes \cite{antonenko2019same} \cite{10.1093/scan/nsy060} \cite{SCHOOT2016454} . Given these findings, IBS has emerged as a promising indicator of communication quality, offering a measurable way to assess the effectiveness of interactions.

Current research indicates that IBS also occur during remote communication, suggesting that brain activities can synchronize between individuals even when they are not physically co-present \cite{SCHWARTZ2022119677} \cite{WIKSTROM2022108316}. However, the level of synchronization during remote interactions is significantly lower compared to face-to-face communication. This attenuation of IBS during technologically-assisted communication, such as video calls, can be attributed to factors like latency and reduced sensory richness, both of which might contribute to the decrease in synchronization \cite{SCHWARTZ2022119677}. Delays inherent in digital communication, such as latency caused by signal processing and transmission, interfere with the rhythm and timing of turn-taking in conversations \cite{boland2022zoom} . These delays can disrupt the natural flow of conversation and make it more challenging for participants to communicate, causing a phenomenon often referred to as ``Zoom fatigue'' \cite{peper2021avoid} \cite{williams2021working}. Consequently, addressing these disruptions and understanding their impact on IBS is essential for enhancing the quality of remote communication. As we delve further into the effects of delays during remote interaction, it becomes clear that mitigating these factors could enhance the level of inter-brain synchronization during remote interactions.

Following the observation that latency inherent in digital communication can disrupt the natural flow of conversation and reduce IBS, this research aims to investigate the extent to which these delays specifically contribute to the reduction of IBS during remote communication. We hypothesize that the presence of latency in remote communication environments leads to a measurable decrease in IBS. To test this hypothesis, we employed an experimental setup that simulates various levels of communication delay, allowing us to systematically assess their impact on IBS. IBS is measured using electroencephalography (EEG) and quantified through Phase Locking Value. Brainwaves are usually classified into different frequency bands, including delta (0.5 to 4 Hz), theta (4 to 8 Hz), alpha (8 to 12 Hz), beta (12 to 35 Hz), and gamma (above 35 Hz), each associated with different cognitive and behavioral states. In this study, we focus on alpha-wave synchronization (8 to 12 Hz) and beta-wave synchronization (13.5 to 29.5 Hz), and analyze the difference between the two frequency bands. Ultimately, this research aims to uncover how communication latency affects synchronization in remote settings and to establish IBS as a reliable indicator of communication quality in such interactions.

\section{RELATED WORK}
\subsection{Inter-brain Synchrony during Communication}
Inter-brain synchronization (IBS) has become a pivotal concept in understanding how neural activities align during communication. This alignment is thought to reflect certain aspects of social interaction, such as interactional synchrony, the anticipation of others' actions, and the co-regulation of turn-taking \cite{10.1371/journal.pone.0012166}. Research has shown various instances where IBS is associated with effective communication, such as in cooperative tasks \cite{10.1093/scan/nsy060} \cite{ShehataENEURO.0133-21.2021}, joint problem-solving \cite{antonenko2019same}, and knowledge sharing \cite{perez2017brain}, where higher synchronization levels often correlate with better comprehension and rapport. For instance, in a state of team flow, where a group is deeply engaged in a task to achieve a common goal, participants showed an increased level of IBS \cite{ShehataENEURO.0133-21.2021}. 

Some other research suggests that synchronization of brain activity can occur even during absence of physical presence. Pérez et al. \cite{perez2017brain} demonstrated that during verbal communication in the same space, brain synchronized even when participants are not able to see each other. Another study compared IBS in real and virtual environments, revealing that participants could achieve synchronization even when interacting through avatars in a virtual reality setting, further supporting the idea that physical presence is not a strict requirement for neural alignment \cite{GUMILAR202162}.

Few IBS research focus on remote communication, where the absence of physical presence, direct non-verbal cues, and communication delays often occur. Schwartz et al. \cite{SCHWARTZ2022119677} found that brain waves still synchronized during remote communication via a video chat tool. However, the study also revealed that the level of synchrony was lower compared to face-to-face conditions, highlighting the challenges in achieving the same degree of neural alignment remotely. 

Extensions of theoretical models, such as the Kuramoto model \cite{kuramoto1975self}, provide insights into how synchronization can persist or break down under delayed coupling conditions \cite{yeung1999time}. These models demonstrate that weakly coupled oscillators can achieve synchronization through phase alignment, even in the presence of delays, up to a critical threshold. This relationship between delays and diminished synchronization mirrors experimental findings, emphasizing the importance of investigating delay-dependent dynamics in IBS. In particular, the study by Adrian P. Burgess \cite{burgess2013interpretation} categorizes synchronization mechanisms into four types: Reciprocal, Induced, Driven, and Coincidental. Among these, the Reciprocal mechanism, characterized by bidirectional coupling, is especially relevant for explaining the IBS during remote communication. Our study further validates this mechanism, showing that synchronization can persist even under delayed conditions. In contrast, the Induced and Driven mechanisms, which rely on external drivers or unidirectional influences, and the Coincidental mechanism, which reflects random alignment, fail to explain IBS during remote communication. Collectively, these findings highlight the pivotal role of Reciprocal coupling in sustaining IBS during remote interactions and underscore the necessity of further investigating how delay-dependent dynamics shape this phenomenon.

\subsection{Inter-brain Synchrony During Different Types of Relationship}
Different types of relationships between individuals can significantly influence the degree of brain activity synchronization during communication. Pan et al. \cite{pan2017cooperation} conducted a study comparing IBS among lovers, friends, and strangers during cooperative tasks. The results indicated that lover dyads exhibited significantly higher synchrony and achieved the best task performance, suggesting that emotional closeness and familiarity enhance neural alignment. In the context of parent-child interactions, a similar phenomenon was observed. Reindl et al. \cite{REINDL2018493} found that significant IBS was present only in parent-child dyads during cooperation, with no such synchronization occurring with strangers. These findings underscore the importance of considering the relationship between dyads when designing experiments, as the nature of the relationship can profoundly affect the outcomes of brain synchronization studies.

\subsection{Latency During Video Conferencing}
Latency in video conferencing tools significantly impacts communication quality and user experience. Latency arises from the various stages of video transmission, including video capture, encoding, decoding, display, and network transmission \cite{6740280}. Boland et al. \cite{boland2022zoom} found that turn-taking delays in Zoom conversations averaged 487 ms, compared to 135 ms for in-person interactions, while technical audio transmission delays in platforms like Zoom typically range from 30 to 70 ms \cite{boland2022zoom}. The effects of latency on communication become particularly noticeable as delays increase. Garg et al. \cite{10.1145/3491101.3519678} observed that delays of 500 ms or more are noticeable during communication. These longer delays can cause various issues, including increased instances of simultaneous speaking, longer periods of mutual silence, and a decreased speaker alternation rate  \cite{boland2022zoom} \cite{SCHOENENBERG2014477}. Additionally, Schoenenberg et al. \cite{SCHOENENBERG2014477} reported that higher latency contributes to more frequent unintended interruptions, further complicating interactions during video conferencing. Understanding and addressing these latency thresholds is crucial, as they directly influence the perceived quality and effectiveness of video-mediated communication.

\subsection{Phase Locking Value}
Phase Locking Value (PLV) is a widely recognized metric in neuroscience research due to its ability to quantify the consistency of phase relationships between brain signals, which is fundamental for exploring functional connectivity in the brain \cite{lachaux1999measuring} \cite{HAKIM2023120354}. Originally introduced by Lachaux et al. \cite{lachaux1999measuring}, PLV provides a robust measure for evaluating the stability of phase differences across repeated trials, making it an invaluable tool in electrophysiological studies to investigate neural synchrony and integration. PLV offers several advantages compared to other methods, including its ability to provide a frequency-specific measure of synchronization, separate phase and amplitude components, and directly quantify transient phase-locking with high temporal resolution, which is particularly useful for studying neural integration during cognitive tasks \cite{lachaux1999measuring}. However, PLV is not without limitations. It is susceptible to volume conduction effects and shared noise sources, which may lead to spurious correlations or hyper-connections, particularly in naturalistic settings where participants share similar sensory experiences \cite{perez2017brain} \cite{burgess2013interpretation} \cite{DIKKER2021117436} \cite{phaselagindex}. Despite these challenges, PLV remains a core tool in hyperscanning research, supporting numerous studies investigating inter-brain coupling and social interactions \cite{HAKIM2023120354}.

\section{METHODS}
\subsection{Participants}

Twenty-two participants (13 males, 9 females) between 21 and 31 years old (M=24.72, SD=2.75) took part in the experiment. They were organized into eleven dyads, with each dyad being acquainted with each other and able to communicate with a same language in native level (7 pairs in Mandarin, 4 pairs in Japanese). All participants had normal vision and hearing, with no known neurological or psychiatric disorders. The participants were university students experienced in using video chat tools. None reported any physical or neurological conditions that would prevent them from participating. Informed consent was obtained from all participants before the experiment. Each participant received a ¥2000 JPY gift voucher as compensation, or ¥3000 JPY if the experiment exceeded two hours. This study was approved by the Ethical Review Board of our university (No. 24-07).

\subsection{Environment Setup}

\begin{figure*}
    \centering
    \includegraphics[width = \textwidth]{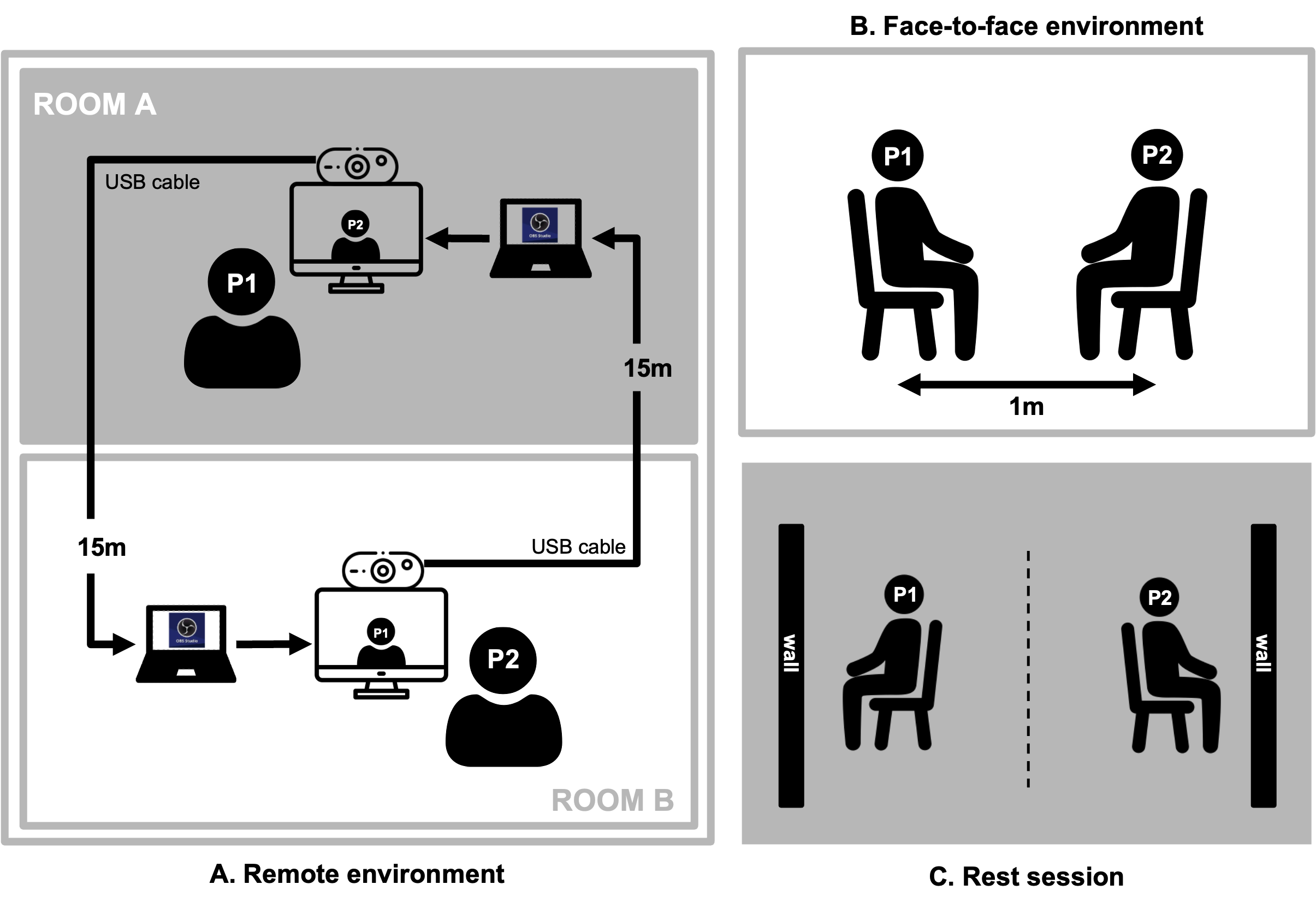}
    \caption[The environment setups.] {\textbf{The environment setups:} (A) Participants communicated from separate rooms in the remote communication setting. (B) Participants communicated face-to-face. (C) During rest sessions, participants did not communicate and could not see each other.}
    \label{setup}
\end{figure*}

\begin{figure}
    \centering
    \includegraphics[width=0.8\linewidth]{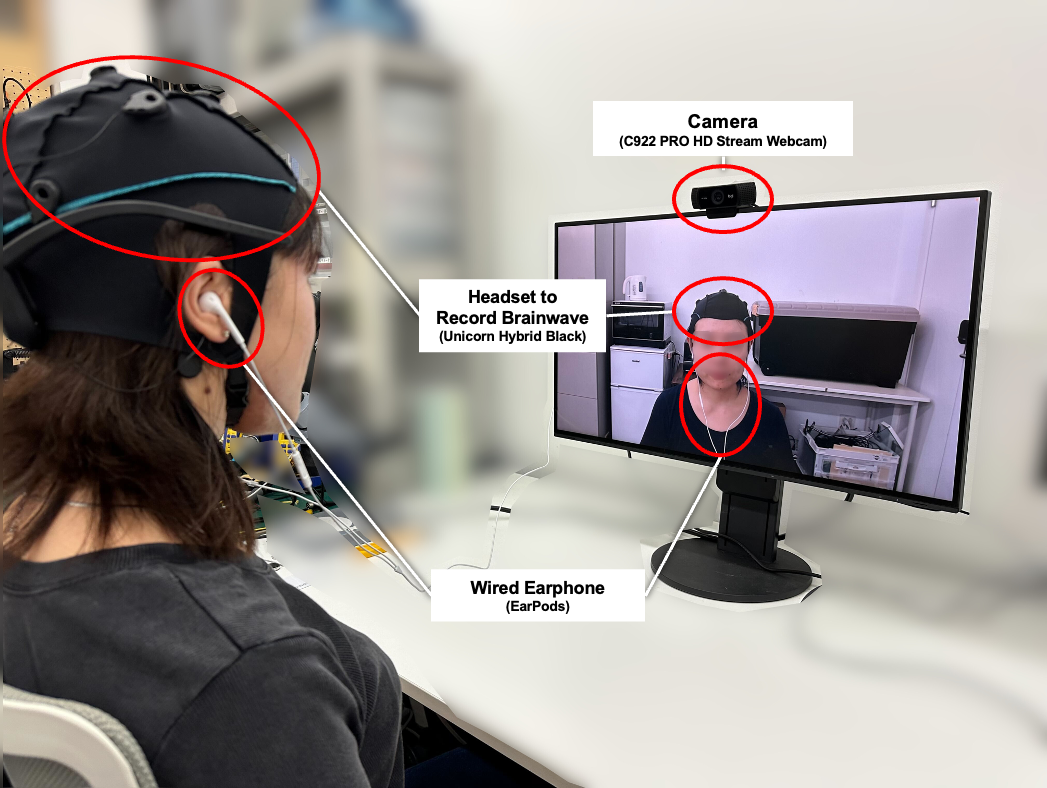}
    \caption{Experiment during remote condition.}
    \label{experiment}
\end{figure}

To conduct a controlled experiment, we established two distinct environments: a face-to-face setting and a remote setting. In the face-to-face environment (Figure~\ref{setup} B), we arranged two chairs facing each other, spaced 1 meter apart. Participants were instructed to sit on these chairs and engage in the experiment while directly facing one another. In the remote environment (Figure~\ref{setup} A), we set two rooms separated by a distance of 12 meters. This distance was chosen to ensure that participants could not hear each other's voices through any means other than the provided headphones at normal speaking volumes. To eliminate the potential for unstable delays caused by network transmission, we connected a laptop running macOS in Room A to a camera (C922 PRO HD Stream Webcam, Logitech International S.A.) in Room B using a 15-meter USB cable. The camera feed was displayed on a monitor (EIZO EV2785-BK) in Room A. Similarly, the monitor in Room B was connected to the camera in Room A through the same method. Each monitor in both rooms was equipped with wired earphones (EarPods, Apple Inc.), enabling participants to communicate with each other during the remote tasks (Figure~\ref{experiment}). The monitors, cameras, and earphones used in both rooms were of the same model to ensure consistency in the experimental setup.

In terms of software, we used OBS~\footnote{OBS, Open Broadcaster Software, \url{https://obsproject.com}} and its plugins to facilitate video calls between participants and to configure the delay settings. Both PCs were set up in the same manner. Since each monitor was connected to the camera in the opposite room, we used OBS to display the video feed from the camera directly on the monitor, allowing participants to see the scene in the other room. We then used OBS's built-in Video Delay plugin to set the video delay, and we configured the audio delay using a separately downloaded VST 2.x Plugin. Both plugins allowed for precise adjustments with a resolution of 1 millisecond. Additionally, to minimize environmental noise, we utilized OBS's built-in Noise Suppression plugin, setting the suppression level to -20 dB.

Although we used the above setup to avoid delays caused by the internet, delays in signal processing were inevitable. To address this, we tested the latency of both the audio and video signals separately. As described in the procedure below, we set delays ranging from 0 to 600 ms in intervals of 150 ms under experimental conditions. Therefore, when confirming the signal processing delay, we also checked the delays in both audio and video under different delay conditions. The results showed that the delays in signal processing remained relatively constant across different delay conditions. The confirmation methods of audio signal and video signal are described as follows.

\begin{figure*}
    \centering
    \includegraphics[width = \textwidth]{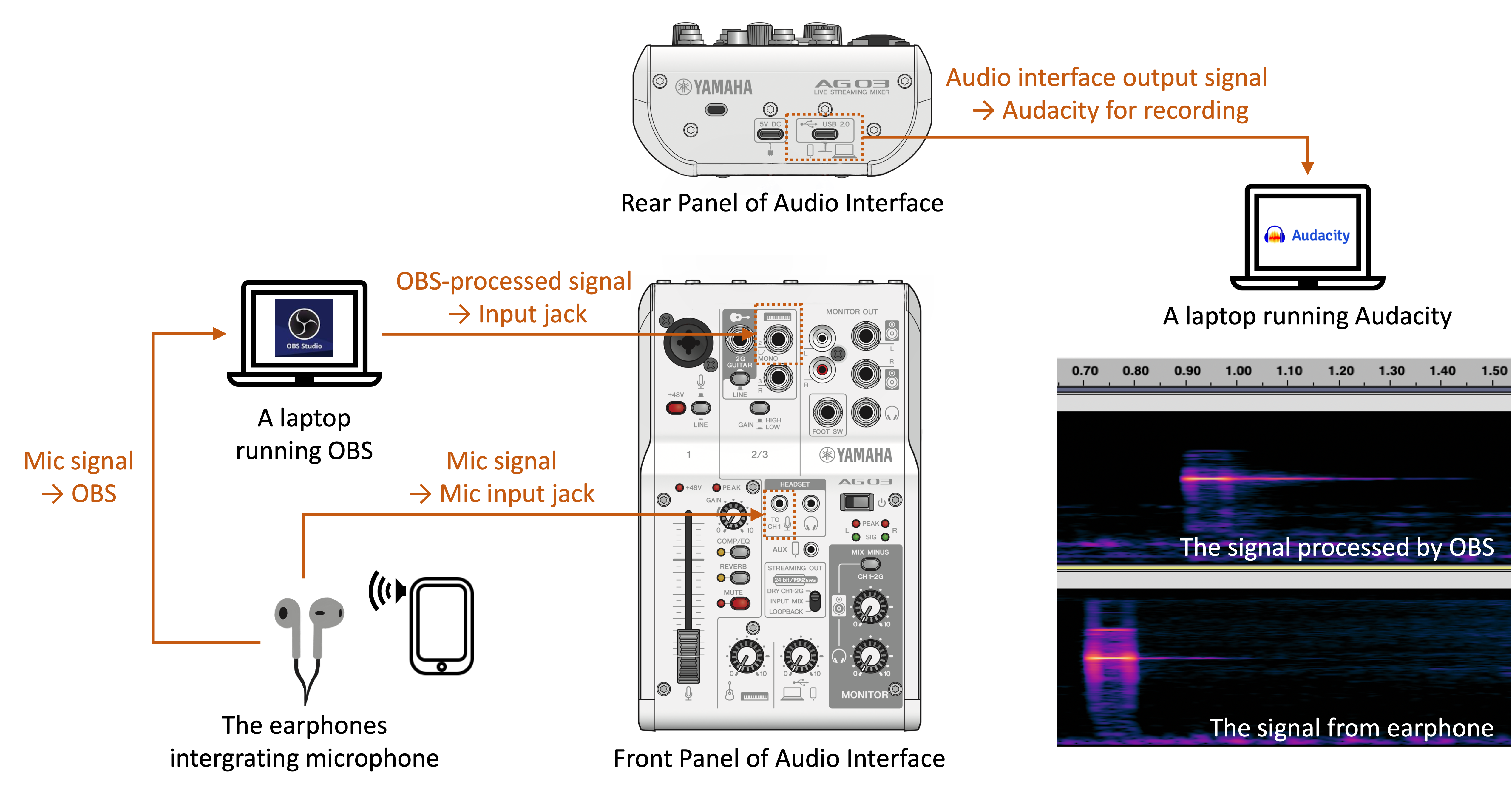}
    \caption{To measure the audio delay in OBS, we play a beep sound from an external device (e.g., a smartphone) and capture it through an audio interface. The captured signal is replayed via OBS, and we calculate the delay between the timestamp of the original beep sound and its output from OBS.}
    \label{delay_audio}
\end{figure*}

\begin{figure*}
    \centering
    \includegraphics[width = \textwidth]{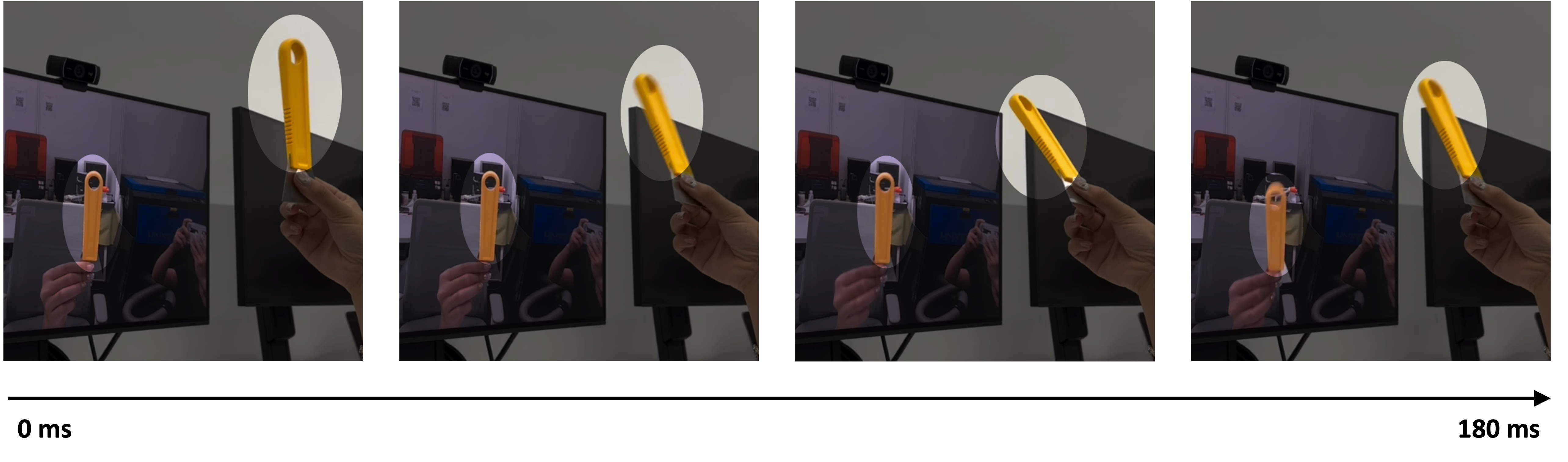}
    \caption{To quantify video delay in OBS, we configure the OBS window to display the camera feed and quickly move a marker in front of the camera. The captured movement is processed and replayed via OBS. To measure the delay, we use a secondary camera to record both the original movement and the OBS playback, then compute the time difference between the timestamps of the original action and its OBS output.}
    \label{delay_video}
\end{figure*}

\begin{itemize}
    \item \textbf{Audio Signal:} As shown in Figure~\ref{delay_audio}, we connected both the earphones' microphone and the laptop's line out to an audio interface (YAMAHA AG03MK2)~\footnote{YAMAHA AG03MK2 user guide, \url{https://jp.yamaha.com/files/download/other_assets/2/1549762/ag06mk2_en_ug_b0.pdf}}. Using a smartphone, we played beep sounds at fixed intervals near the earphones' microphone. In this setup, we could hear both the original sound from the phone (representing the person speaking in a conversation) and the sound processed by the OBS software (representing the sound heard by the other person in a conversation). We recorded both sounds and then analyzed them with the audio processing software named Audacity~\footnote{Audacity, \url{https://www.audacityteam.org}}. The results showed that the sound processed by OBS had a delay of approximately 180 ms compared to the original sound. 
    \item \textbf{Video Signal:} As shown in Figure~\ref{delay_video}, we connected a laptop to an external monitor and placed a web camera at the center of the top edge of the monitor, which was also connected to the laptop. Through OBS, the video feed from the web camera was displayed on the screen. A test participant then moved a reference object in front of the camera, while another participant simultaneously recorded both the participant and the video processed by OBS on the screen from a third-person perspective using a 60fps camera. Using video editing software, we confirmed that the video processed by OBS had approximately a 180 ms delay compared to the original video.
    
\end{itemize}

In addition to the conversation task, we also included a rest session (Figure~\ref{setup} C) designed to capture EEG data during periods of silence. During rest sessions, participants did not communicate and could not see each other. Rest sessions were conducted in both environmental settings: face-to-face setting and remote setting. In the face-to-face setting, they faced different blank walls in the same room, while in the remote setting, they were physically separated into different rooms facing a non-displaying monitor. EEG data in the rest session was collected after the participants had fully relaxed.

\subsection{Procedure}
\subsubsection{Calibration}
To obtain the highest quality EEG data, participants wear EEG caps that are carefully selected to match their head size, ensuring a snug fit that fully covers the scalp and maintains optimal contact with the skull. Once the appropriate cap is selected, electrodes are installed, and conductive gel is applied. After the gel application, we use Unicorn Suite software to verify the quality of the EEG signals. If any bad channels are detected, we address them by reapplying gel and adjusting the electrodes as needed until all channels show high-quality output.

\subsubsection{Experiment}
The entire experiment consisted of eight tasks, divided into two main sessions: face-to-face and remote. In the remote session, delays for video and audio were set simultaneously in five different tasks, with delays set to 0 ms, 150 ms, 300 ms, 450 ms, and 600 ms. An experimental procedure without considering the random sequence is shown in Table~\ref{tab:experiments}, with each task lasting 3 minutes. Each group of participants experienced the tasks in a randomized order. The randomization included both the sequence of the sessions and the order of tasks within each session. However, to ensure participants were fully relaxed before the rest task, it was always placed at the beginning or the end of the two main sessions.

In all tasks except the rest tasks, participants engaged in conversational tasks, where they were asked to discuss a specific Aesop's Fable. At the beginning of each task, participants were provided with a prepared fable to read. They were then instructed to discuss the story during the conversation. After each task, participants were asked to remove their earphones, and were individually asked whether they experienced any discomfort during the conversation.

Finally, after the experiment, both participants returned to the same room, where we explained the delays that occurred during the tasks and asked if they noticed the delays or perceived differences in the delay between each task.

\begin{table}
    \centering
    \caption{The experimental procedure without considering the random sequence.}
    \begin{tabular}{| c | c | c |}
        \hline
         \textbf{Session} & \textbf{Task} & \textbf{Delay(ms)} \\
         \hline
         \multirow{2}{*}{\textbf{Face-to-face}}& conversation & -\\  \cline{2-3}
         & rest & - \\  \hline
         \multirow{6}{*}{\textbf{Remote}} & conversation & 0 \\ \cline{2-3}
         & conversation & 150 \\ \cline{2-3}
         & conversation & 300 \\ \cline{2-3} 
         & conversation & 450 \\ \cline{2-3}
         & conversation & 600 \\ \cline{2-3}
         & rest & - \\ \hline
    \end{tabular}
    \label{tab:experiments}
\end{table}

\subsection{EEG Hyperscanning Recordings}
\begin{figure}
  \centering
    \centering
    \includegraphics[width=0.8\linewidth]{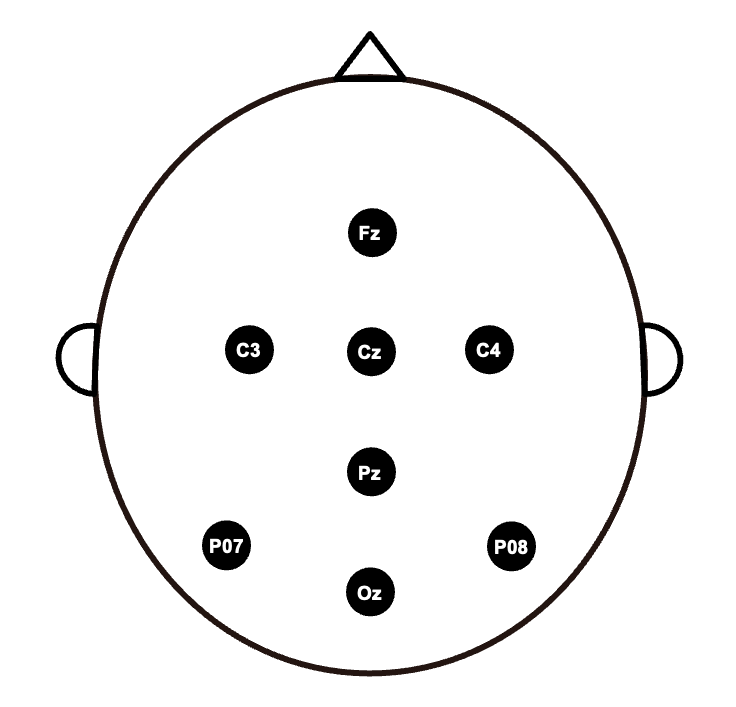} 
    \caption[The position of electrodes] {\textbf{The position of electrodes:} The device has 8 electrodes evenly distributed on the scalp.}
    \label{electrodes}
\end{figure}

EEG data were recorded simultaneously using two Unicorn Hybrid Black headsets (g.tec Medical Engineering, Austria), each equipped with 8 electrodes positioned at Fz, C3, Cz, C4, Pz, PO7, Oz, and PO8 according to the 10/20 system (Figure~\ref{electrodes}). The electrode positions remained unmodified to ensure even scalp coverage and a balanced representation of neural activity. To achieve precise temporal alignment between the headsets, we synchronized the EEG signals using the Lab Streaming Layer (LSL) protocol. The signals were collected at a sampling rate at 250 Hz. To ensure high-quality data, the headsets were used in the wet condition (with gel), although they support both dry and wet configurations.

\subsection{EEG Data Analysis}
\subsubsection{Preprocessing}
The EEG data were first preprocessed using the EEGLAB toolbox (v2024.0) in MATLAB (version 2023b, The MathWorks Inc.), along with custom scripts. Initially, a 2 to 30 Hz band-pass filter was applied to the recorded signals. The EEG data for each dyad were then split into separate files for each participant. Next, from the original 3-minute recordings, 2 minutes of data were selected for each trial, with a focus on segments that had the fewest channel rejections as identified by the PREP pipeline \cite{bigdely2015prep}. The PREP pipeline was also used to detect bad channels, which were then interpolated. After that, the data were segmented into 5-second epochs without overlapping. Finally, an adaptive mixture independent component analysis (AMICA) was performed on each participant's data  to decompose the EEG signals into independent components for artifact rejection. Artifacts are automatically removed by enabling component rejection, which runs up to 15 iterations, checking every 1 iteration to reject any component whose likelihood deviates by more than 3 standard deviations. This setup systematically identifies and removes components that deviate significantly from typical neural signals, thereby filtering out artifacts. 

\subsubsection{Synchronization Analysis}
Inter-brain synchrony was quantified from EEG signals of both users using the phase-locking value (PLV), which is a widely used measure in EEG data analysis to assess the synchronization between different brain regions. In this study, all eight EEG channels from each participant were used, resulting in 64 channel pairs for the analysis. The PLV calculation method, rooted in the approach detailed by Lachaux et al. \cite{lachaux1999measuring}, provides a robust measure of phase synchronization, crucial for understanding the dynamics of brain connectivity during cognitive tasks. In this approach, PLV is defined as follows:
\[
    \text{PLV} = \frac{1}{N} |\sum_{n=1}^{N} e^{j\theta(t, n)}|
\]
where:
\begin{itemize}
    \item $\theta(t, n) = \phi_1(t, n) - \phi_2(t, n)$ represents the phase difference between two signals at time point $t$ during $n$ trial.
    \item $\phi_1(t, n)$ and $\phi_2(t, n)$ are the instantaneous phases of the two signals from participants in the same dyad.
    \item $j$ is the imaginary unit ($j = \sqrt{-1}$).
\end{itemize}
To compute PLV for the data, the EEG signals were first filtered within two frequency bands: the alpha band (8 to 12 Hz) and the beta frequency band (13.5 to 29.5 Hz). This filtering was achieved using a finite impulse response (FIR) filter and the Hilbert transform was applied after filtering. The PLV was then calculated by determining the phase differences between 64 pairs of EEG signals (8 channels $\times$ 8 channels) across trials. This approach ensured comprehensive analysis of synchronization across all channel pairs. The resulting PLV values ranged from 0 to 1, where a value of 1 signifies perfect phase synchrony between the signals, and 0 indicates no synchrony. High PLV values suggest a strong functional connection between brain regions, indicative of coherent neural communication, while lower values point to a lack of synchronization, reflecting potential disruptions in neural interaction. This process resulted in PLV values that reflect the consistency of phase synchronization between brain regions, facilitating the analysis of functional connectivity during the experimental conditions.

\subsubsection{Statistical Analysis}
To evaluate PLV, we compared the PLV values from the actual data to those derived from a surrogate data. The surrogate data were generated by pairing participants of the same dyad at non-time-corresponding moments during each trial. The epochs of the EEG data for participant B were shuffled so that the order of the epochs was different from that of participant A. For example, the EEG signal of participant A during the first epoch was paired with the EEG signal from the ninth epoch, a non-corresponding epoch, of participant B. To assess the statistical significance of these comparisons, we employed a permutation test, which is ideal for EEG data analysis as it does not require normality assumptions \cite{good2013permutation}. We conducted 10,000 permutations for each test, generating a distribution of possible outcomes by reshuffling the data multiple times. This comparison against surrogate data, consistent with previous two-brain research \cite{perez2017brain} \cite{SCHWARTZ2022119677}, allowed us to assess the extent to which the observed neural coupling reflected true inter-brain synchrony rather than coincidental alignment.

In addition, for the channel‐pair analysis, we employed the same permutation test procedure described above to compare each pair’s PLV values from the true data against those from surrogate data. However, given that we tested 36 possible channel pairs, the resulting p‐values were Bonferroni‐corrected to address the issue of multiple comparisons. This ensures that any observed differences in inter‐brain synchrony are not inflated by chance findings across the numerous electrode pairs.

To compare IBS between face-to-face communication and remote communication under specific delay conditions (0 ms, 150 ms, 300 ms, 450 ms, and 600 ms), we used a Kruskal-Wallis test followed by multiple comparisons. This approach allowed us to determine if the differences between face-to-face communication and each delay condition were statistically significant, providing a more detailed understanding of how delays in remote communication affect inter-brain synchronization.

\section{RESULTS}

\begin{table*}
    \centering
    \caption{The result comparing actual and surrogate data for alpha and beta bands.}
    \renewcommand{\arraystretch}{1.2}
    \begin{tabular}{| c | c | c | c | c | c |}
        \hline
         \multirow{2}{*}{\textbf{Session}} & \multirow{2}{*}{\textbf{Task}} & 
         \multicolumn{2}{c|}{\textbf{Alpha Band}} & \multicolumn{2}{c|}{\textbf{Beta Band}} \\
         \cline{3-6}
         & & \textbf{Average PLV} & \textbf{P-value} & \textbf{Average PLV} & \textbf{P-value} \\
         \hline
         \multirow{2}{*}{Face-to-face}
         & Conversation & 0.02612 & 0.0001 & 0.02563 & 0.0001 \\ \cline{2-6}
         & Rest & 0.02391 & 1.0000 & 0.02444 & 0.9053 \\ \hline
         \multirow{6}{*}{Remote} 
         & 0 ms & 0.02565 & 0.0001 & 0.02558 & 0.0001 \\ \cline{2-6}
         & 150 ms & 0.02532 & 0.0005 & 0.02530 & 0.0002 \\ \cline{2-6}
         & 300 ms & 0.02526 & 0.0002 & 0.02548 & 0.0001 \\ \cline{2-6}
         & 450 ms & 0.02387 & 0.9815 & 0.02421 & 0.9996 \\ \cline{2-6}
         & 600 ms & 0.02420 & 0.9999 & 0.02411 & 0.7141 \\ \cline{2-6}
         & Rest & 0.02386 & 1.0000 & 0.02429 & 1.0000 \\ \hline
    \end{tabular}
    \label{tab:result}
\end{table*}

\begin{figure*}
  \centering
    \centering
    \includegraphics[width = \textwidth]{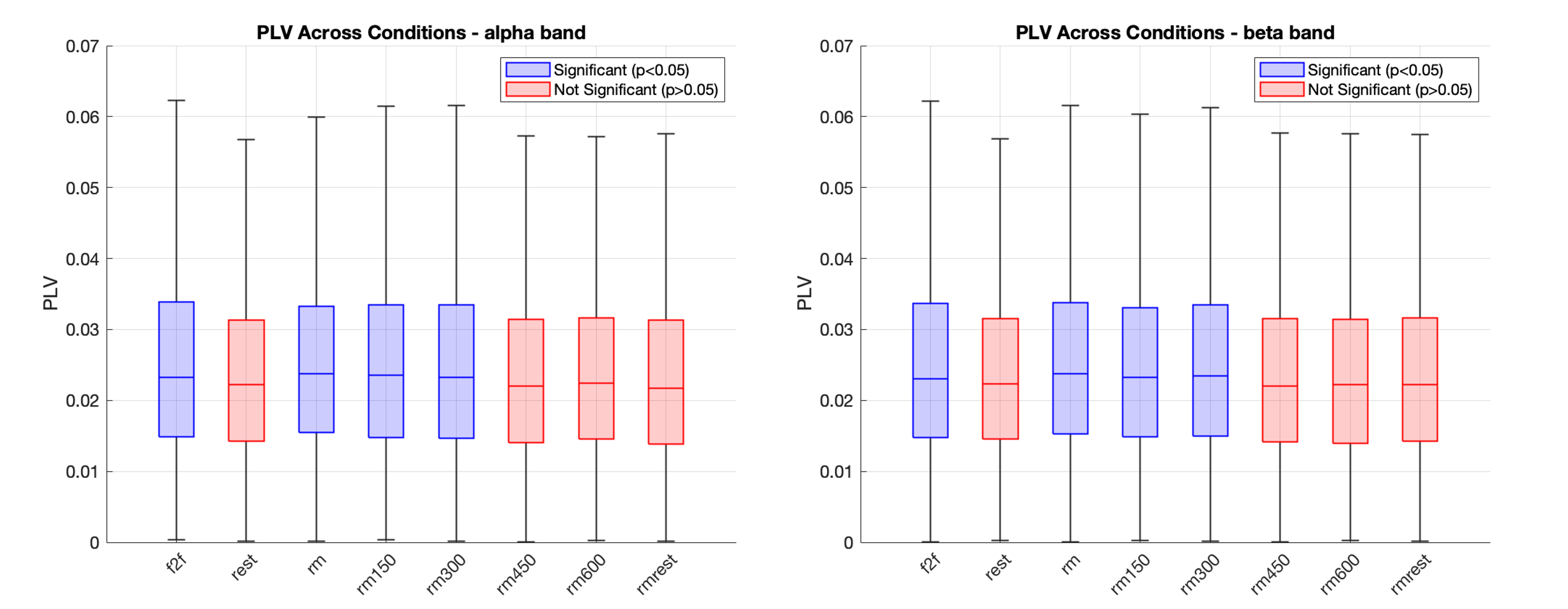}
    \caption{\textbf{PLV Comparison Across Conditions in Alpha and Beta Bands:} Phase-Locking Value (PLV) across all conditions for the Alpha Band (left) and Beta Band (right). The boxplots represent PLV values for each condition, with the colors indicating statistical significance. Blue boxplots denote conditions where PLV is significantly different from the surrogate data ($p<0.05$), while red boxplots represent conditions where the difference is not statistically significant ($p>0.05$). Conditions include face-to-face (f2f), rest, and various remote communication delay settings (rm, rm150, rm300, rm450, rm600, and rmrest).}
    \label{plv}
\end{figure*}

\subsection{Comparison of Actual and Surrogate Data for Inter-Brain Synchrony}
Differences between the actual and surrogate data were analyzed using permutation tests across all conditions---including face-to-face communication, remote communication with varying delays, and rest conditions in both environmental settings (Table~\ref{tab:result}, Figure~\ref{plv}). Here, the independent variable was the communication condition (e.g., face-to-face vs. remote with a specific delay), and the dependent variable was the PLV measure of inter-brain synchrony. The results for both the alpha and beta bands showed similar patterns. In both rest conditions, regardless of the environment (face-to-face or remote), the result of the permutation test was not significant ($p > 0.05$). Conversely, we observed significant inter-brain synchrony during some communication conditions. Specifically, in face-to-face communication, p-value was lower than 0.05, indicating stronger neural synchronization. A similar trend was seen in remote communication with no delay (0 ms) and delays of 150 ms and 300 ms ($p < 0.05$). However, differences were not significant ($p > 0.05$) at higher degrees of delays, 450 ms and 600 ms. These findings demonstrate that inter-brain synchronization in both the alpha and beta bands is preserved under shorter delays (0 ms to 300 ms) but significantly disrupted under longer delays (450 ms and 600 ms), supporting the hypothesis that communication delays critically affect inter-brain synchrony.

To contextualize these findings, we compared the observed PLV values with those reported in prior research. The beta band PLV values during face-to-face communication and remote communication without delay ranged from 0.0178 to 0.0338, representing the average PLV across all channel pairs (across dyads). In prior research by Pérez et al. \cite{perez2017brain}, beta band PLVs were reported to range from 0.0311 to 0.0393. While the results in this study fall slightly below this range, they overlap at the lower end. This overlap suggests that the inter-brain synchronization observed in face-to-face communication in the study aligns with established patterns. The slight differences may reflect variations in experimental setups, participant characteristics, or task conditions. This comparison further validates the observed neural coupling as meaningful and consistent with prior research.

\subsection{Inter-brain Synchrony Across Experimental Conditions}

\begin{figure*}
  \centering
    \centering
    \includegraphics[width = \textwidth]{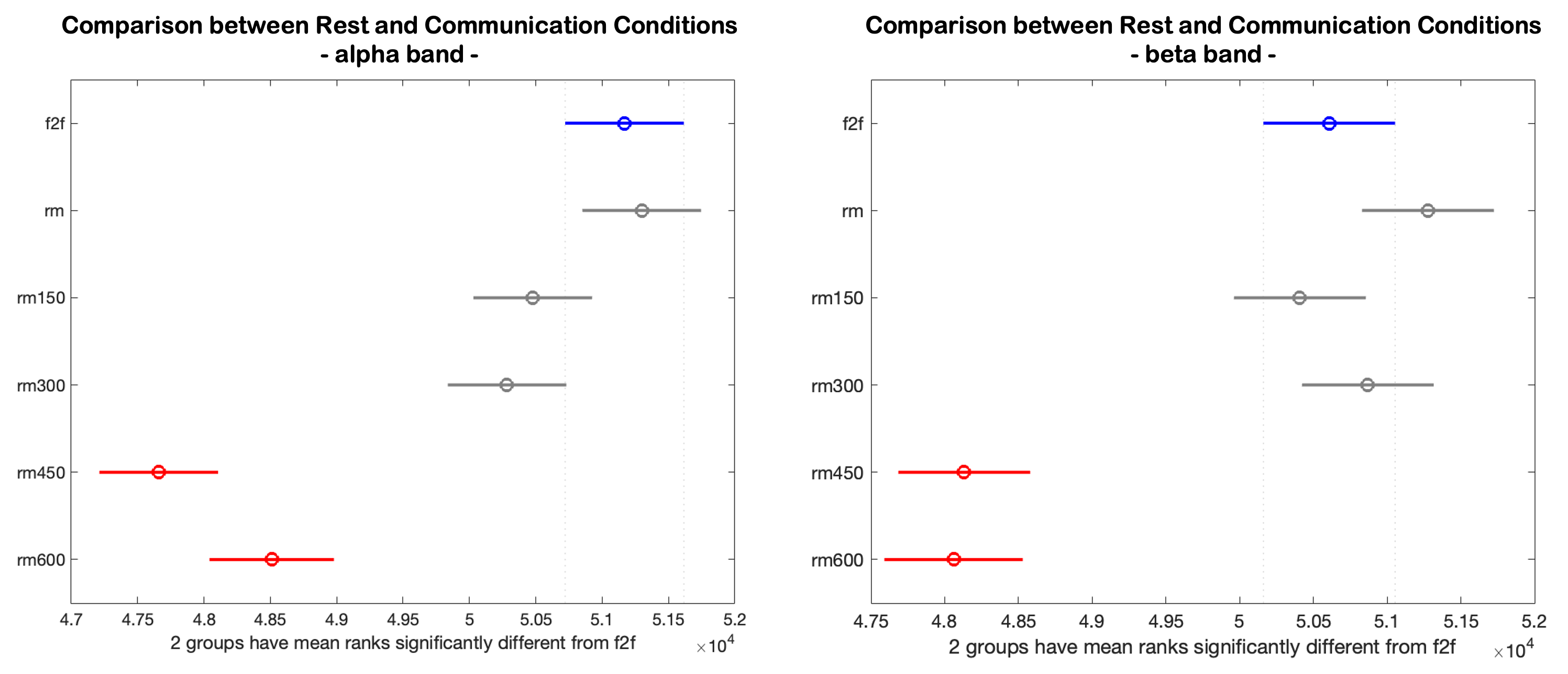}
    \caption{PLV was significantly lower than face-to-face at delays over 450 ms (red) but remained comparable at delays under 300 ms for remote communication (gray). The x-axis of this figure shows the mean rank of each condition based on the results of the Kruskal-Wallis test to evaluate the statistical difference between each condition.}
    \label{compf2f}
\end{figure*}

We conducted a Kruskal-Wallis test to compare face-to-face communication with remote communication across different delay conditions. In this analysis, the independent variable and dependent variable were also communication condition and the PLV value, respectively. For both the alpha and beta frequency bands, the analysis revealed a significant overall effect of communication condition on inter-brain synchronization ($\chi^2(5) = 220.08$, $p < 0.05$ for alpha; $\chi^2(5) = 199.19$, $p < 0.05$ for beta).

Following the Kruskal-Wallis test, multiple comparisons were performed to determine which specific conditions differed from face-to-face communication (Figure~\ref{compf2f}). The results for both frequency bands indicated that the 450 ms and 600 ms delay conditions showed significantly lower levels of neural synchronization compared to face-to-face communication ($p < 0.05$). In contrast, the 0 ms, 150 ms, and 300 ms conditions did not show significant differences from face-to-face communication ($p > 0.05$), suggesting that shorter delays in remote communication did not disrupt inter-brain synchronization to a significant degree. These results in both frequency bands suggest that while shorter delays in remote communication (0 ms to 300 ms) allow for levels of neural synchrony comparable to face-to-face communication, higher delays (450 ms and 600 ms) significantly disrupt synchronization, supporting the hypothesis regarding the impact of transmission delays on IBS.

\subsection{Inter-brain Synchrony Across Channel Pairs}

\begin{figure*}
  \centering
    \centering
    \includegraphics[width = \textwidth]{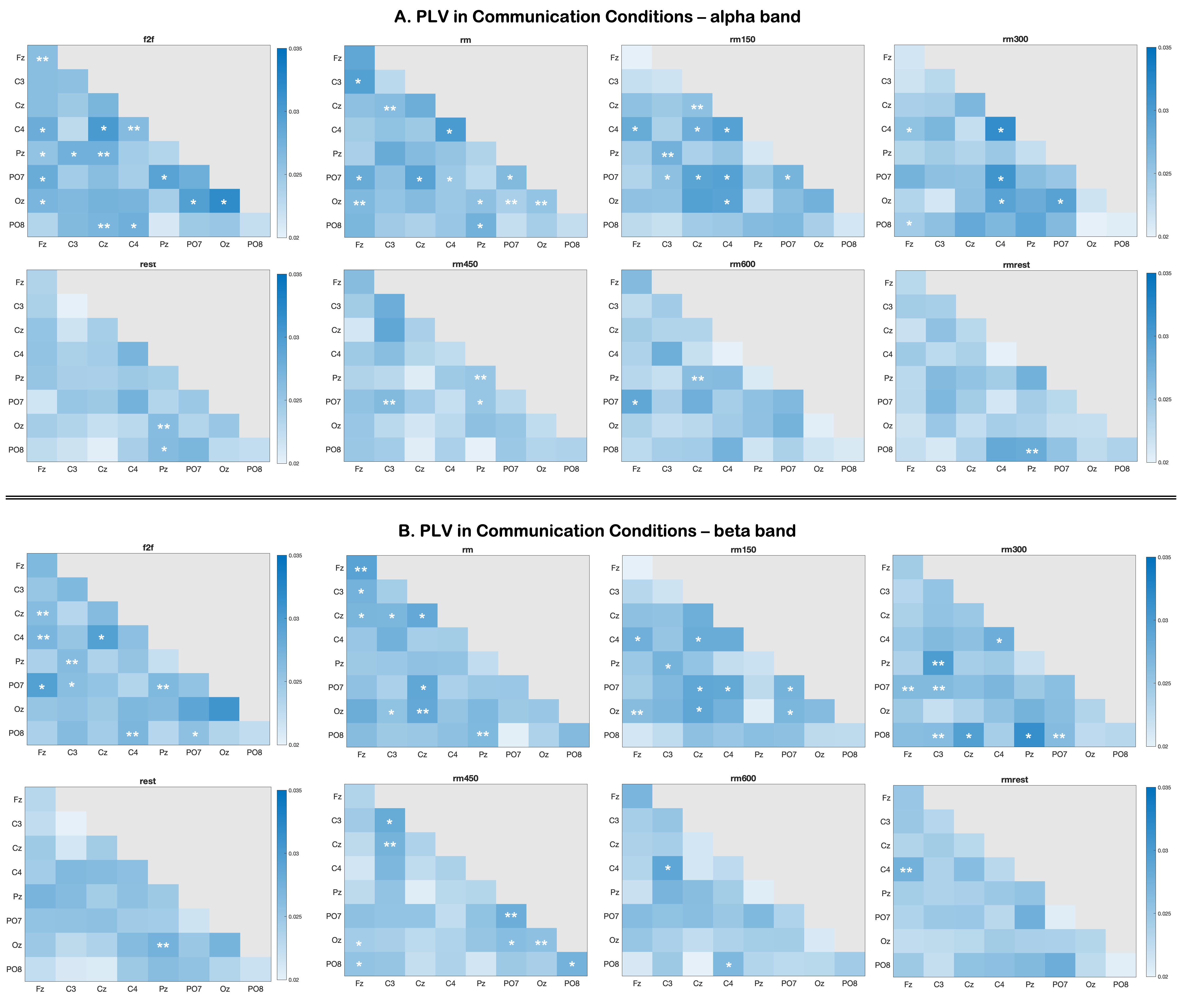}
    \caption{\textbf{Inter‐Brain Synchrony in Alpha and Beta Bands Under Communication:} Pairwise PLV‐based inter‐brain synchrony matrices for (A) the alpha frequency band (8–12 Hz) and (B) the beta frequency band (13.5–29.5 Hz) across eight tasks: face‐to‐face (f2f), remote (rm), remote with delays of 150 ms (rm150), 300 ms (rm300), 450 ms (rm450), 600 ms (rm600), and two resting‐state tasks in face‐to‐face (rest) and remote (rmrest) settings. The color scale represents PLV magnitude, with darker shades indicating stronger synchronization. Asterisks (*, **) mark statistically significant channel pairs, where * corresponds to $p< 0.005$ and ** corresponds to $p< 0.05$. Each matrix shows the inter‐brain coupling across electrodes Fz, C3, Cz, C4, Pz, PO7, Oz, and PO8.}
    \label{allchannels}
\end{figure*}

\subsubsection{Alpha band}
In the alpha band (Figure~\ref{allchannels} A), face‐to‐face communication conditions consistently yield broader, highly significant synchronization across the brain. By contrast, remote conditions consistently showed lower or less‐widespread alpha coupling, especially at the frontal electrode (Fz). Within the remote conditions, however, we can distinguish two clusters: (1) rm, rm150, rm300, where central and posterior electrodes (Cz, C4, Pz, PO7) still exhibited moderately strong and significant alpha synchronization, and (2) rm450, rm600, which tended to show a further reduction in significant inter‐brain synchrony. Although parietal‐occipital coupling remained relatively intact under remote conditions, these results suggest that longer delays can further interrupt or weaken frontal and midline alpha‐band synchronization.

\subsubsection{Beta band}
In the beta band (Figure~\ref{allchannels} B), face‐to‐face communication also shows a comparable level of synchronization, with multiple electrode pairs—from frontal‐central to parietal‐occipital—often reaching very low p‐values ($p < 0.05$). However, the distribution of significant beta links in the face‐to‐face condition tends to be more localized than in the alpha band. In the remote conditions, although beta synchronization remains in the same general numerical range, many pairs lose their significance or exhibit less‐widespread coupling as the delay rises. Still, remote conditions, with delays ranging from 0 ms to 300 ms, preserve pockets of robust beta synchronization, especially at central and posterior electrodes (Cz, C3, PO7), while longer delays (rm450, rm600) show a further weakening of beta coherence, particularly at frontal‐central sites.

\subsection{Interview}
The study included two types of interviews: one conducted after each task during remote communication and another conducted after the entire experiment. In both interviews, we focused on participants' experiences with latency.

\subsubsection{After Task}
During the post-task interviews, eight out of twenty-two participants reported experiencing latency during the high-delay conditions. Three participants mentioned noticing delays specifically in the 600 ms condition, while the other five participants reported delays in both the 450 ms and 600 ms conditions. Participants described unintended interruptions caused by the delay, which made it difficult to anticipate when the other person was about to speak. For example, Participant 8 noted, ``There's a little delay, our words conflicted while having conversation,'' while Participant 16 mentioned, ``I can feel there is a delay when another participant heard my response,'' both in the 600 ms delay condition. These comments illustrate how artificial delays disrupted the natural flow of conversation during high-latency conditions.

\subsubsection{After the Entire Experiment Process}
In the post-experiment interviews, four participants among the fourteen participants who did not initially report feeling delays during the post-task interviews acknowledged that, while they did notice delays, they perceived them as a normal aspect of remote communication and did not find them disruptive enough to mention at the time. Participant 3 stated, ``I thought it was normal to have delays during remote communication.'' On the other hand, the remaining participants stated that they genuinely did not perceive any delay during the trials. 

Overall, nearly half of the participants (n = 10) did not notice latency during communication, while some considered the delays inherent to remote communication. This divergence in experiences suggests that participants have varying sensitivities to communication delays.

In addition to latency, participants also reported other discomforts during remote communication, such as difficulty in making eye contact and the lack of body language. Participant 6 remarked, ``It's difficult to make eye contact because of the different position of the camera and monitor,'' while Participant 13 said, ``Compared to face-to-face communication, I feel there is less body language.''

\section{DISCUSSION}
This study investigated the impact of transmission delays on IBS during remote communication and compared it with face-to-face communication, using the PLV as the primary metric. The results revealed that while IBS is present in both face-to-face and low-delay ($\leq$ 300 ms) remote conditions, it becomes significantly disrupted when delays exceed 450 ms. Interestingly, PLV values for low-delay remote conditions were similar to face-to-face interactions, while rest conditions in both environments showed similar low PLV values. Additionally, interviews with participants revealed divergent experiences with communication delays, with some perceiving them as inherent and non-disruptive, while others found them noticeable and intrusive.

The comparison between actual and surrogate data for inter-brain synchrony indicates that neural synchronization was significant during face-to-face communication and low-delay remote communication ($\leq$ 300 ms) in both the alpha and beta bands. However, the result of permutation test in the high-delay condition ($\geq$ 450 ms) highlights the critical threshold at which transmission delay begins to disrupt neural coupling. This finding reinforces the notion that transmission delays—especially those beyond 450 ms—have a detrimental impact on the quality of communication by impairing neural synchrony, which has been shown to be essential for effective interaction and engagement.

The comparison of PLV values across conditions reinforces these findings. The low synchrony observed during rest in both face-to-face and remote settings indicates that physical proximity alone does not enhance synchronization without active engagement. This highlights that inter-brain synchrony relies on interaction. Notably, the similarity between face-to-face and low-delay remote conditions in both the alpha and beta bands suggests that remote communication can emulate the natural flow of in-person interactions in terms of brainwave synchronization when delays are minimal. This aligns with previous research showing that face-to-face communication \cite{antonenko2019same} \cite{10.1093/scan/nsy060} \cite{SCHOOT2016454} and remote communication \cite{SCHWARTZ2022119677} both foster neural synchrony , while providing new evidence that delays under 300 ms allow remote interactions to sustain comparable synchrony. Conversely, the significant differences between high-delay conditions ($\geq$ 450 ms) and face-to-face communication emphasize the disruptive effect of substantial delays, impairing the natural rhythm of interaction, reducing neural synchrony, and potentially lowering communication quality.

When comparing alpha‐band and beta‐band inter‐brain synchronization under varying communication delays, several important observations emerge. First, alpha‐band synchronization consistently appears broader and more stable, exhibiting consistently higher PLV values, especially during face-to-face communication and remote conditions without delays. Interestingly, in the remote condition with 300 ms and 450 ms delays, the beta‐band plots reveal a higher number of significantly synchronized channels compared to their alpha‐band counterparts, suggesting that beta‐range activity may remain relatively robust under moderate to mid‐range delays. By contrast, alpha‐band synchronization, though generally more pervasive at lower delays, appears more susceptible to attenuation at these same delay levels. Given these findings, beta‐range measures might be more appropriate for analyzing inter‐brain synchrony in remote communication scenarios where moderate temporal offsets are unavoidable, whereas alpha‐range measures might be more informative in setups with minimal or no delay.

Interviews revealed varying sensitivities to communication delays among participants. Over half of the participants (n = 12) noticed latency, with 8 finding it disruptive and interrupted the natural flow of conversation, while others were unaffected or unaware of the delays. These differences may stem from factors like familiarity with technology, task context, or individual cognitive processing. Interestingly, while IBS was found to be disrupted when the delay exceeded 450 ms, some participants only noticed the delay in the 600 ms condition. This suggests that inter-brain synchrony might be disturbed even when participants do not consciously perceive or consider the delay. This finding highlights the potential of IBS as an objective measure of communication quality that could detect subtle disruptions not always perceived by users.

Overall, these results highlight how transmission delays can disrupt neural synchrony, sometimes well before participants become consciously aware of any latency. By revealing distinct alpha- and beta-band patterns under varying latency conditions, this study underscores the sensitivity of IBS—particularly when delays exceed 450 ms—as a meaningful metric for assessing remote communication effectiveness. Importantly, IBS has the potential to identify not only the impact of delay but also the key factors that contribute to the quality and effectiveness of remote communication more broadly.

\section{LIMITATIONS AND FUTURE WORK}
This study presents several limitations that should be addressed in future research. First, the sample size was relatively small, which may limit the generalizability of the findings, though the statistical analyses conducted were significant. Additionally, we did not analyze synchronization differences at the individual EEG channel level, which may have uncovered more nuanced patterns of intra-brain synchrony.

Another limitation lies in the controlled nature of the delay used in this experiment. While this study focused on fixed transmission delays, real-world remote communication typically involves fluctuating latencies due to network instability. These sudden changes in delay, which were not simulated here, could impact IBS differently. Future research should explore more dynamic, real-world scenarios to better understand how unpredictable latency affects inter-brain synchronization.

Furthermore, the study did not account for other factors influencing IBS in remote communication. Previous research has identified elements contributing to discomfort during video conferencing, such as excessive eye gaze and reduced mobility \cite{Bailenson2021Nonverbal}. Interviews in this study also revealed discomfort from unnatural eye contact. These factors are likely to affect neural synchronization and should be examined in future research.

Finally, future work could focus on developing strategies to address disruptions in IBS during remote communication. For example, systems could provide real-time feedback to users about their synchronization levels or adapt dynamically to maintain better synchronization under varying conditions. Additionally, since multimodal interaction influences social presence and cognitive load, future research should examine how differences in sensory richness (e.g., audio-only vs. video-audio communication) affect neural coupling in remote settings. Furthermore, individual differences in neural plasticity and communication styles may also impact IBS, potentially limiting the generalizability of our findings. Future studies could explore how experience with remote collaboration, cognitive flexibility, and cultural variations in communication strategies shape neural synchrony. By leveraging IBS as a measurable indicator of communication quality, these approaches could help minimize the negative effects of high latency and enhance the overall effectiveness of remote collaboration.

\section{CONCLUSION}
In conclusion, this study contributes to the growing body of research on inter-brain synchronization and remote communication by demonstrating that IBS can occur during remote communication but is highly dependent on transmission delay. Transmission delays of more than 450 ms significantly disrupt synchrony, whereas low-delay remote communication can achieve levels of neural synchronization comparable to face-to-face interactions. These findings highlight the potential of using IBS as a tool to assess communication quality in remote settings and suggest that reducing transmission delays should be a key focus in the development of remote communication technologies.

\begin{acks}
This work was supported by JST Moonshot R\&D Grant JPMJMS2012, and JPNP23025 commissioned by the New Energy and Industrial Technology Development Organization (NEDO).
\end{acks}

\bibliographystyle{ACM-Reference-Format}
\bibliography{sample-base}

\end{document}